\documentclass[a4paper,11pt]{article}
\usepackage[utf8]{inputenc}
\usepackage[T1]{fontenc}
\usepackage{graphicx,url}
\usepackage{a4wide}
\usepackage{mathpazo}
\usepackage[font=footnotesize,labelfont=bf]{caption}
\usepackage{amsmath}
\usepackage{bm}
\usepackage{amssymb}
\usepackage{multirow}
\usepackage{color,soul}
\usepackage{xurl}
\usepackage{authblk}
\usepackage[hidelinks]{hyperref}
\usepackage{adjustbox}
\usepackage[shortlabels]{enumitem}
\setlist[enumerate]{nosep}
\usepackage[table,xcdraw]{xcolor}
\usepackage{tabulary}
\usepackage{booktabs}
\usepackage{lineno}
\usepackage{float} 

\usepackage{verbatim}
\newcommand{\detailtexcount}[1]{%
  \immediate\write18{texcount -merge -sum -q main.tex > main.wcdetail }%
  \verbatiminput{main.wcdetail}%
}

\title{Women’s mobility networks enable more efficient travel}

\author[1]{Sílvia de Sojo}
\author[1,2]{Sune Lehmann}
\author[1]{Laura Alessandretti}
\affil[1]{DTU Compute, Technical University of Denmark}
\affil[2]{Center for Social Data Science, University of Copenhagen}

\begin{document}


\maketitle

\begin{abstract}
Our understanding of gender differences in mobility is marked by a clear tension: surveys portray women’s movements as more complex than men's, while digital traces suggest less diverse travel.
Here, we resolve the contradiction by modeling trajectories as networks of sequential visits, using smartphone traces linked to self-reported gender for 543,155 individuals across 10 countries.
We show that the apparent conflict in the literature arises because women’s mobility networks are simultaneously more clustered and more home-anchored—a nuance obscured by aggregate metrics.
This pattern arises because women tend to link multiple destinations within single trips—for trips spanning up to 150 km and multiple days. 
This organization yields systematically higher travel efficiency, measured as distance saved through destination chaining over monthly sequences.

\end{abstract}


\section*{Introduction}
The importance of studying urban mobility from a gendered perspective has been widely recognized across the fields of geography, transportation, and social sciences~\cite{rosenbloom_trip-chaining_1989, law_beyond_1999, schwanen_how_2008, uteng_gendered_2008, singh_is_2020, hayati_sustainable_2020, delmelle_women_2025}
and, more recently, by quantitative studies in Human Mobility research~\cite{gauvin_gender_2020, alessandretti_scales_2020, collins_spatiotemporal_2024, gauvin_gaps_2024, althoff_large-scale_2017}, facilitated by the widespread adoption of smartphones and wearable devices~\cite{brockmann_scaling_2006, gonzalez_understanding_2008, song_modelling_2010, pappalardo_returners_2015, alessandretti_evidence_2018, schlapfer_universal_2021, kraemer_mapping_2020}. 
There is broad agreement that there are indeed gender differences within mobility behaviors: women tend to travel shorter distances~\cite{gauvin_gender_2020, lenormand_influence_2015, yuan_analyzing_2016, reisch_behavioral_2021, battiston_revealing_2022, xian_gender_2025} and engage in less physical activity than men~\cite{althoff_large-scale_2017}.  

Yet, when it comes to the patterns of how and when locations are visited---how people organize their movements across space---there are contradictory findings in the literature.
Research based on travel surveys often describes women’s location visit patterns as more complex and diverse than men’s~\cite{hayati_sustainable_2020, duchene_gender_2011, scheiner_womens_2017, susilo_changes_2019}.
This is typically attributed to women visiting a broader range of locations for a wider variety of purposes and relying on more flexible modes of transport~\cite{law_beyond_1999, hayati_sustainable_2020, hanson_impact_1981, ng_understanding_2018}.
Women are also frequently observed to combine multiple visits within a single trip ---a pattern commonly linked with their disproportionate engagement in caregiving and domestic responsibilities~\cite{rosenbloom_trip-chaining_1989, duchene_gender_2011, de_madariaga_women_2013, loukaitou-sideris_gendered_2016, scheiner_womens_2017, susilo_changes_2019}.
Findings based on large-scale mobile phone datasets, however, point in the opposite direction.
With few exceptions~\cite{psylla_role_2017}, studies report that women visit fewer locations~\cite{gauvin_gender_2020, contreras_linking_2023}, display less diverse location visits~\cite{gauvin_gender_2020, yuan_analyzing_2016, contreras_linking_2023, macedo_differences_2022, centellegher_job_2025, bertocchi_big_2025}, and are as predictable in their mobility patterns as men~\cite{song_limits_2010}. 

In this work, we resolve the disagreement between survey-based and big-data views of gendered mobility by showing that the structural organization of movements provides a critical explanatory dimension. 
Leveraging a 10-country smartphone dataset of 543,155 individuals, including self-reported gender, we represent each trajectory as a mobility network, in which links encode empirical sequences of visited locations.

This dataset allows us to quantitatively identify novel gender differences in the structure of location-visit patterns that conventional aggregate metrics overlook.
We show that structural differences reflect women’s consistent tendency to travel more efficiently across short and long distances.

\section*{Results}
A mobility trajectory is an ordered list of the places a person visits.
In this study, we focus on one-month subsequences of each person’s trajectory (see Methods:~\nameref{sec:meth-netw}) and transform every subsequence into an individual mobility network. 
Nodes correspond to visited locations, and an edge connects two nodes whenever those locations are visited consecutively. 
Edge weights capture the number of transitions. 

Representing mobility as a network is both intuitive and analytically powerful. 
It is intuitive because networks mirrors how people naturally conceptualize space, as places connected by familiar paths~\cite{lynch_image_1960}. 
The analytical power lies in the ability to capture diverse aspects of mobility within a single formalism. 

Individual mobility networks were introduced in earlier works~\cite{song_limits_2010, schneider_unravelling_2013, rinzivillo_purpose_2014}, but prior studies have largely focused on lower-dimensional descriptors---such as recurrent network motifs in daily mobility~\cite{schneider_unravelling_2013} or activity types inferred from trajectories~\cite{rinzivillo_purpose_2014}---leaving much of the rich behavioral structure they encode largely unexplored.

Here, we analyze male and female mobility networks using measures of increasing complexity. 
We begin with basic network descriptors---the nodes and edge weights---and then examine how locations are interconnected through sequential visits.
Gender is self-reported as a binary (a limitation of the dataset) and we use \textit{male/man} and \textit{female/woman} interchangeably, as no universal standard exists~\cite{hawkesworth_sex_2013, ritz_we_2024} (see Methods:~\nameref{sec:meth-data}).

\subsection*{Men are more active and have more heterogeneous behavior}\label{sec:Nk}
 
In the mobility network representation, the number of nodes, $k$, and the sum of edge weights, $N$, capture two fundamental aspects, with $k$ reflecting an individual’s number of unique locations visited--hereafter their \emph{repertoire size}-- and $N$ the total number of visits within a month--hereafter their \emph{activity} (see Methods:~\nameref{sec:meth-netw}).

\begin{figure}[H]\centering\includegraphics[width=1\linewidth]{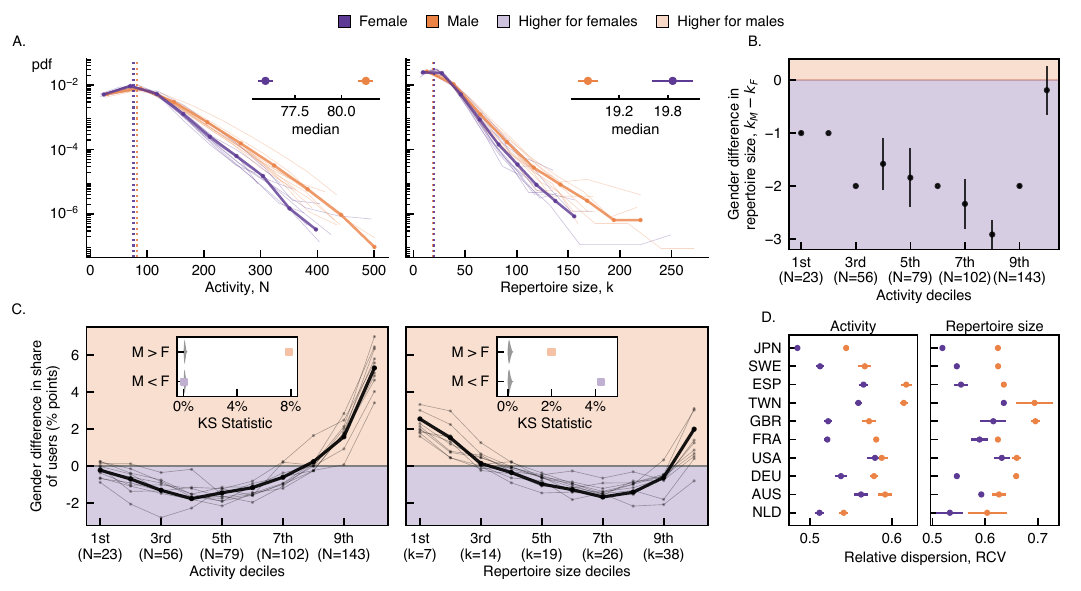}
    \caption{\textbf{Women explore more locations with fewer visits; men dominate the extremes.} 
        \textbf{a.} Distributions of activity and repertoire size for females (purple) and males (orange). Dashed lines mark bootstrapped medians; thin lines show country-level distributions; the bold line shows country-balanced results. Insets display the mean and standard error of the bootstrapped medians (see Methods).
        \textbf{b.} Gender differences in median repertoire size by activity decile. Error bars indicate the standard error of the difference in medians, estimated via country-balanced bootstrapping (Methods). Background shading highlights where the repertoire is larger for males (light orange) or females (light purple).
        \textbf{c.} Difference between the share of male and female users across deciles of activity (right) and repertoire size (left), shown separately by country (thin lines) and country-balanced sample (bold line, Methods). 
        Insets report Kolmogorov–Smirnov (KS) statistics for the hypotheses M > F (light orange) and M < F (light purple), compared against random baselines with permuted gender labels (gray violin plots).
        \textbf{d.} Robust coefficient of variation (RCV, see Methods) for females (purple) and males (orange) by country. Error bars show the standard error of the bootstrapped RCV. 
}
    \label{fig:1}
\end{figure}

We begin by assessing differences across genders in a standard way: comparing median values.
Typically, according to this measure, men appear more active, with a median monthly visit count of $81.31\pm0.40$ (vs. $75.91\pm0.39$ for women), while women visit slightly more unique locations ($19.86\pm0.25$ vs. $18.82\pm0.12$ for men, see Fig.\ref{fig:1}a inset. 
For details on the median bootstrap, see Methods:~\nameref{sec:meth-imbalance}).
This pattern is remarkably robust when we organize users according to their activity, with women consistently visiting more unique locations than men across most activity deciles (Fig.\ref{fig:1}b), a result that holds across countries (see Supplementary Material, SM~\ref{sisec:bsctry}).

Comparing median values alone, however, does not capture the full richness of the gendered behavioral differences because the distributions of $N$ and $k$ are heavy-tailed (Fig.\ref{fig:1}a). 
When we examine the gender distribution across repertoire size deciles, we find that men are overrepresented at both the upper and lower extremes in most countries (Fig.\ref{fig:1}c, right).
Hence, the number of locations is not simply systematically higher for females. Rather, there is a broader variability for males.
We confirm the significance of these results with one-sided Kolmogorov–Smirnov (KS) tests (see Methods:~\nameref{sec:meth-KSstat}, Fig.\ref{fig:1}c inset, and results across countries in SM~\ref{sisec-KS}),
and quantify the broader dispersion of males' distribution with the robust coefficient of variation (RCV) across counties (Methods:~\nameref{sec:meth-rcv}; Fig.\ref{fig:1}d).

As shown in the results above, there are substantial heterogeneities across individuals (see Fig.\ref{fig:1}).
As we move toward analyzing richer aspects of mobility encoded in individual networks, direct comparisons across individuals with vastly different repertoire sizes (nodes) or activity would be misleading. 
To account for this, we stratify individuals based on their activity and repertoire size. 
Given the strong correlation between $N$ and $k$, ($\rho_{N,k} = 0.72$), we group individuals into three subgroups based on their combined deciles: Inactive (1st–3rd deciles), Moderately Active (4th–7th deciles), and Active (8th decile and above).
Men are overrepresented in the Inactive and Active groups, while women are more often Moderately Active (see SM~\ref{sisec-Nkstrat}).
Since $35.06\%$ of users fall outside these subgroups, we also consider the complete sample in our analysis.

\subsection*{Structural differences in gendered mobility networks}\label{sec:struct}
Building on basic mobility features---activity and repertoire---we now analyze mobility sequences through network-level metrics to uncover structural gender differences.

Our analysis focuses on three key aspects: 
(i) the average \emph{clustering coefficient}, capturing the connectivity, or tendency of visited locations to form tightly connected subgroups;
(ii) \emph{number of cycles}, capturing the occurrence of specific visit sequences; and
(iii) \emph{degree centrality of the top three nodes},  reflecting the extent to which the most connected locations link to others. 
These widely used measures provide a concise and complementary characterization of network structure.
More details on these metrics are provided in Methods:~\nameref{sec:meth-nwmet}. 

To isolate structural network differences from those driven by activity and repertoire, we use nearest-neighbor matching to find pairs of similar males and females based on the number of visits and distinct locations visited~\cite{stuart_matching_2010}.
Then, for each metric under study $X$, we focus on the average difference between male and female pairs relative to their mean ($\frac{X_{Male} - X_{Female}}{(X_{Male} + X_{Female})/2}$), and compare these results to a randomized experiment where gender labels are permuted (see Methods:~\nameref{sec:meth-matching}).

\begin{figure}[H]
    \centering
    \includegraphics[width=1\linewidth]{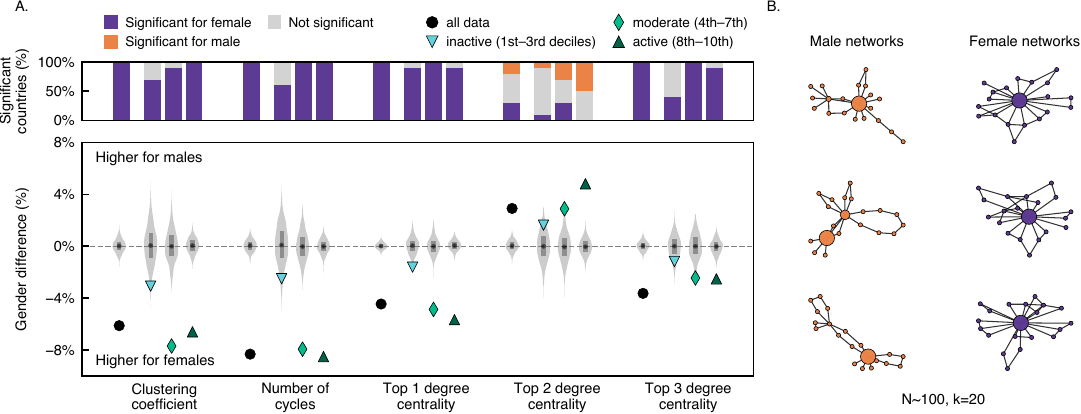}
    \caption{
    \textbf{Women's networks are more interconnected, with home as a central anchor}
    \textbf{a.} (bottom) Average relative difference across male-female pairs matched by activity and repertoire size (see Methods)for different network metrics (from left to right): clustering, number of cycles, and degree of the first, second, and third highest-degree nodes. 
    For each metric, markers represent different activity groups (see legend). Gray violins represent the reference distribution obtained from 1,000 random shuffles of gender labels. 
    (top) The fraction of countries in which each metric is significantly larger for females (purple bars), males (orange bars), or not significant (gray bars). Significance is assigned when the observed difference exceeds the 95th percentile of the shuffled distribution in the corresponding direction.
    \textbf{b.} Example networks for moderately active males (orange) and females (purple). Nodes represent locations, links indicate sequential visits, and node size reflects visit frequency (scaled exponentially to enhance visibility). Distances between nodes are proportional to their geographical separation.
    The example networks were selected to illustrate the contrasts shown in subplot (a), with the following criteria: female networks with high clustering and top-ranked degree centrality, and male networks with low clustering and high second-ranked degree centrality.    }
    \label{fig:2}
\end{figure}

We find pronounced structural differences between males' and females' mobility networks. 
On average, women’s networks are characterized by higher \emph{clustering} than men’s (see Fig.\ref{fig:2}a bottom), suggesting that women have a tendency to connect different locations rather than frequently repeating the same trip sequences.
This result is supported by other connectivity metrics---including density, transitivity, and diameter---detailed in SM~\ref{sisec:match-extnwmet}.

Additionally, we find that, on average, women have more \emph{cycles} in their mobility networks, reflecting a tendency to visit multiple locations within a single trip---a behavior usually referred to as trip-chaining~\cite{rosenbloom_trip-chaining_1989, primerano_defining_2008}.

Finally, moving to nodes' \emph{centrality}, we find that the top node by number of connections---typically home~\cite{schneider_unravelling_2013,rinzivillo_purpose_2014} (see SM~\ref{sisec:1degect-Home} and Methods:~\nameref{sec:meth-howde})---has, on average, a higher degree for women than for men, indicative of a more pronounced home-anchoring behavior.
Similarly, the third-most-connected node tends to be more prominent in women’s networks than in men’s.
In contrast, the second-most-connected node shows higher centrality among men. 
This node is often associated with the workplace (for 54\% of users in our dataset; see SM~\ref{sisec:1degect-Home}).

Strikingly, these structural gender differences remain overall robust across activity levels (see Fig.~\ref{fig:2}a bottom), and countries (see Fig.\ref{fig:2}a top, SM~\ref{sisec:match-nwmetCtry}).
Additionally, KS tests reveal that, for most metrics, the differences in average behavior reported above arise from a systematic shift of the distribution of one gender relative to the other rather than from changes in the shape or variability of the distributions (see SM~\ref{sisec:KS-nwmet}).

These findings demonstrate consistent structural gender differences: women’s networks are denser and richer in cycles, and are more strongly anchored to the home node (see Fig.~\ref{fig:2}b).

\subsection*{The structure of tours shapes gendered mobility networks}
The higher clustering and prevalence of cycles in women’s mobility networks raise the question of what drives these structural differences. 
We address this by analyzing how men and women chain visits to different locations.

\begin{figure}[H]
    \centering
    \includegraphics[width=1\linewidth]{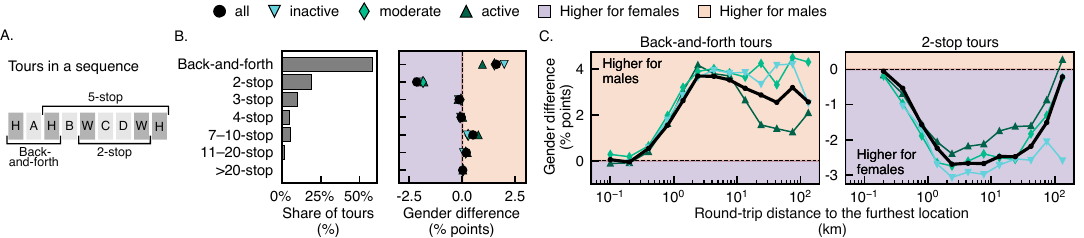} 
    \caption{\textbf{Women make more two-stop tours, men favor back-and-forth}
        \textbf{a.} Identification of tours in a monthly sequence of visits. 
        The rectangles illustrate a visit sequence, with each distinct location represented by a different letter. 
        Anchor locations are shown in dark gray; other visited locations in light gray (see Methods).
        \textbf{b.} Distribution of tours by length (right) and their corresponding differences between the share of male and female tours (left). Marker colors denote activity groups: inactive (light blue downward-pointing triangles), moderately active (green diamonds), active (dark green upward-pointing triangles), and all individuals (black circles). Results show means and bootstrapped standard errors averaged across countries; error bars are too small to be visible (see Methods).
        \textbf{c.} Gender differences (as defined in panel b) in the prevalence of back-and-forth tours (top) and two-stop tours (bottom), grouped by round-trip distance to the furthest location. Lines show country-averaged results for activity groups, with marker colors as in panel b.
}
    \label{fig:jrny-nav}
\end{figure}

We turn our attention to individuals' \textit{tours}: sequences of stops that begin and end at the same anchor location (see Fig.\ref{fig:jrny-nav}a, and Methods:~\nameref{sec:meth-tours})~\cite{daisy_trip_2020, paleti_joint_2011, primerano_defining_2008}.

We find that most tours follow a \emph{back-and-forth} structure (e.g., A–B–A), accounting for 58\% of cases (see Fig.\ref{fig:jrny-nav}b left). \emph{Two-stop tours}---those that chain two stops before returning to the origin location (e.g., A–B–C–A)--- make up 19\%.
Notably, women take two-stop tours more often than men, while men tend to rely on back-and-forth travel (see Fig.\ref{fig:jrny-nav}b right), a pattern robust across countries (see SM~\ref{sisec:tourlen-ctry}). 
For longer tours, gender differences are minimal, except for trips with 7–10 stops (covering $\sim$6\% of all tours), which are more dominant in men. 
We note that not all tours form closed cycles: the sequence of stops A–B–C–D–A creates a loop, whereas the sequence A–B–C–B–A retraces the route in reverse---forming a palindromic pattern. 
Men show a higher share of palindromes ($17.25\pm0.02\%$ for men vs. $12.98\pm0.02\%$ for women), indicating a greater tendency to retrace steps even for longer tours.

The finding that women make more two-stop tours than men is consistent with previous evidence that women are more likely to link several activities into a single tour—often when running errands or meeting daily obligations near home—a pattern known as trip chaining~\cite{rosenbloom_trip-chaining_1989, duchene_gender_2011, de_madariaga_women_2013, loukaitou-sideris_gendered_2016, scheiner_womens_2017, susilo_changes_2019}.
However, we find that the tendency to connect stops extends beyond patterns of day-to-day mobility. 

Grouping tours by the round-trip distance to the furthest location---defined as twice the distance from the anchor location to the most distant stop in the tour---shows that this pattern holds for tours across all distance levels up to 150 km, with the largest disparity observed at 2–5 km (Fig.\ref{fig:jrny-nav}c).
This gendered behavior is observed consistently across countries, and also when grouping tours by duration (see SM~\ref{sisec:tourlen-dist-ctry} and~\ref{sisec:tourlen-durat-ctry}).
Hence, the greater prevalence of two-stop structures among women cannot be explained by trip-chaining to nearby locations alone, but instead points to intrinsic differences in how trips are organized.

\subsection*{Women travel more efficiently}\label{sec:eff}
Concatenating stops into a single tour can improve travel efficiency by avoiding repeated returns to a starting point, such as home. We now assess whether the greater connectivity of women’s mobility networks translates into more efficient travel by considering the spatial embedding of locations.

\begin{figure}[h!]
    \centering
  \includegraphics[width=1\linewidth]{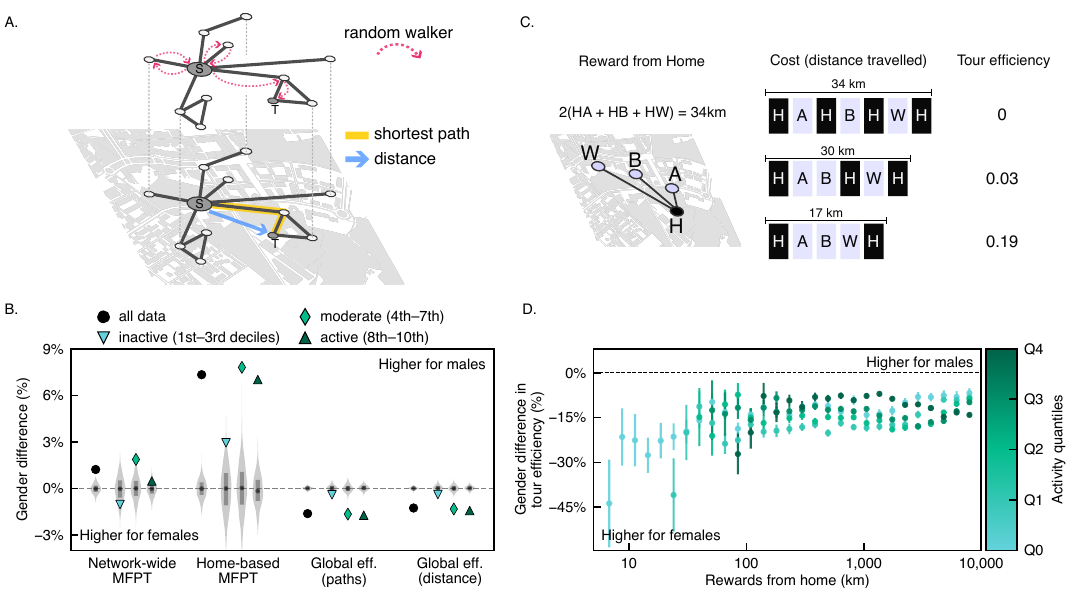}
    \caption{\textbf{Women’s visit patterns support more efficient mobility.}
        \textbf{a.} Schematic of two routing models from a source to a target node: random diffusion (top network, pink dashed line) and shortest path (bottom network, yellow line). The blue arrow depicts the physical distance between source and target.
        \textbf{b.} Average relative differences across male-female pairs matched by activity and repertoire size (see Methods) in network-wide and home-based mean first passage time (MFPT), global efficiency (unweighted and weighted by distance).
         Marker colors denote activity groups. Gray violins show reference distributions from 1,000 random shuffles of gender labels.
        \textbf{c.} Minimal illustration of tour efficiency, defined as the proportion of distance saved when sequencing visits instead of making independent round-trips from home (marked as H in black). Chaining nearby stops reduces the total travel cost while achieving the same overall reward.
        \textbf{d.} Relative gender difference between the average male and female tour efficiency against tour reward (see panel c). Results are shown for different activity quantiles (colorbar).  
        The tour efficiency is computed as $1-\frac{\text{cost}}{\text{reward}}$ for each monthly sequence. 
        Error bars show the standard error of the bootstrapped difference.
}
    \label{fig:eff}
\end{figure}

We begin by formalizing what we mean by `efficiency' in terms of networks, intuitively conceptualized as the average ease of movement between any two nodes in a particular mobility network.
We characterize networks through two routing models: random diffusion and shortest-path navigation (to isolate differences unrelated to activity or repertoire, we resort to nearest-neighbor matching and report the average difference between male and female pairs relative to their mean; see Methods).
The diffusion model captures how easy it is to get somewhere without intent and not knowing the network, while the shortest path model measures how easy it is to navigate a network with intent and perfect knowledge of the network layout (see Fig.\ref{fig:eff}A; Methods:~\nameref{sec:meth-nwnavig}).
We expect real-world behavior to lie somewhere in-between.

First, we simulate a random walker and compute two random-diffusion metrics: the \textit{network-wide mean first passage time} (MFPT), defined as the average number of steps needed to reach a target node for the first time across all source–target pairs, and the \textit{home-based MFPT}, which restricts the target to the home node~\cite{lovasz_random_1993, newman_networks_2016} (see Methods:~\nameref{sec:meth-mfpt}).

Female networks show, on average, shorter \emph{network-wide MFPTs}, except among the most active users (Fig~\ref{fig:eff}B).
This effect is particularly pronounced for trips ending at the home node, \emph{home-based MFPT}, reflecting higher navigability and a stronger home-centered structure.

Second, we measure the \emph{global efficiency}~\cite{latora_efficient_2001, vragovic_efficiency_2005} by comparing shortest path lengths between all node pairs to those in a fully connected network (Fig.\ref{fig:eff}A bottom network).
Across all but the least active group, female networks are, on average, more efficient, requiring fewer intermediate stops to reach any destination.
Weighting shortest paths by log-transformed physical distances shows that women’s networks connect locations with not only fewer intermediate stops but also shorter travel distances (Fig.\ref{fig:eff}B).
Results are overall consistent across countries, with few country- and activity-level exceptions (see SM~\ref{sisec:navmet-ctry}).

Having established that women's networks enable more efficient travel, we now turn to realized behavior. 
Random walks and global efficiency estimate the efficiency attainable under typical routing strategies across all origin–destination pairs, yet actual travel patterns differ in both the locations visited and the strategies employed.
To capture this behavior, we study tour efficiency of monthly, time-ordered sequences of visits.

We define \textit{tour efficiency} as the proportion of distance saved: the difference between a one-month sequence’s estimated \textit{reward}---approximated as the sum of log-transformed round-trip distances from home to each location---and its \textit{cost}---the log-transformed total distance traveled---normalized by the \textit{reward}.
When the \emph{tour efficiency} tends to 1, it indicates that the sequence's cost is minimal relative to visiting each location independently, while a value of 0 indicates that no distance is saved by the traveler through sequencing (see minimal example in Fig.\ref{fig:eff}C). 

This definition relies on three key choices.
First, we construct sequences at the monthly scale to capture how travel is organized across activities that are coordinated over multiple days. 
Second, consistent with the literature, our definition assumes that more distant locations offer greater reward, reflecting the notion that travelers are willing to travel farther only for more valuable destinations~\cite{stouffer_intervening_1940}.
Finally, we measure both reward and cost using log-transformed travel distances rather than travel time.
While travel time is considered a more direct proxy for perceived effort, estimating it reliably would require non-sparse trajectories, city-specific travel conditions, and assumptions about transportation modes, which are not feasible in our setting. 
Log-transformed distance provides a reasonable proxy, given its established power-law relationship with travel time~\cite{barbosa_human_2018}, and the strong empirical correlation between monthly traveled distance and travel time ($\rho_{\text{time,distance}}=0.74$; see SM~\ref{sisec:travel-time}).

Comparing individuals with similar activity levels and sequence rewards, we find that women consistently have higher tour efficiency, typically by 10–20\% relative to males (see Fig.\ref{fig:eff}d). 
This implies that, considering males and females with a similar cost and activity level (proxy for total travel time), females consistently achieve higher reward (see SM~\ref{sisec:cost-reward}). 
This pattern is robust across countries (see SM~\ref{sisec:eff-ctry}).
Taken together, these results suggest that women's mobility is optimized for connectivity and accessibility, potentially reflecting distinct travel needs and behavioral strategies compared to men.

\section*{Conclusion}
In this work, we used large-scale smartphone mobility data from over half a million individuals across ten countries to uncover structural gender differences in mobility networks.
In line with previous findings, we found that males are typically more active compared to females. 
We showed, however, that males are also characterized by greater heterogeneity, both in terms of the number of visits and the number of unique locations visited. 
This finding highlights the limitation of relying solely on measures of central tendency, which can obscure important distributional differences.
Focusing on structural patterns, we found that women’s mobility networks are consistently denser, contain more cycles, and are more anchored to home than men’s, even when controlling for differences in activity and distinct locations visited.
These structural differences are robust across countries and are reflected in tour organization: women more often connect multiple stops within a single trip, and this pattern holds for both everyday travel and less frequent, long-distance trips. 
Embedding networks in geographic space, we showed that these properties of women’s networks enable more efficient travel and that women’s realized travel sequences exhibit higher efficiency across both short and long distances.
These differences in efficiency may reflect both gendered mobility constraints—such as greater caregiving responsibilities~\cite{rosenbloom_trip-chaining_1989, de_madariaga_women_2013} and greater reliance on public transport~\cite{duchene_gender_2011}—as well as more structured coordination of activities~\cite{law_beyond_1999, scheiner_womens_2017, hayati_sustainable_2020}.

Taken together, our findings help understand differences between the data-driven and survey-based perspectives on gendered mobility.
The greater clustering of females’ mobility networks aligns with survey-based evidence that women exhibit more complex and diverse mobility patterns. 
Conversely, women's stronger home anchoring is consistent with digital-trace studies showing that women’s visits are less diverse.

We note a number of limitations to this work.
Our analysis relies on self-reported binary gender, which does not capture the full spectrum of gender identities and roles.
Furthermore, we do not directly observe trip purposes, socioeconomic factors, or constraints such as caregiving responsibilities, all of which are likely to contribute to the observed patterns.
In addition, our efficiency measure is based on travel distance rather than travel time, which may imperfectly capture perceived travel costs; nevertheless, analyses on a subset of high-quality trajectories indicate a strong correspondence between distance and travel time.
Finally, smartphone-derived location data may be affected by sampling biases in device ownership, usage patterns, and data coverage~\cite{barreras_exciting_2024}.

By demonstrating that structural organization---rather than simple diversity metrics---can reveal substantial and systematic gender differences, this work highlights the need for mobility studies to adopt richer network-based approaches.
In doing so, it strengthens the link between large-scale computational studies and decades of evidence from transport and geography research, opening new possibilities for integrating behavioral theory with empirical studies based on large-scale data.
Such integration is key to inform transport planning, urban design, and policy interventions aimed at addressing gendered inequalities in mobility.

\section*{Methods}
\subsection*{Data description}\label{sec:meth-data}
This dataset uses mobility data collected between 2017 and 2019, previously analyzed in~\cite{alessandretti_scales_2020}, where stop locations were identified using the scalable stop-location detection algorithm InfoStop~\cite{aslak_infostop_2020}. 
The stop location data captures the mobility patterns of 543,155 users.
All users included had at least four months of data with positions recorded on at least 80\% of days. To ensure equal representation, we analyzed monthly sequences and sampled three months per user.

Individuals are located across ten countries (Japan, the United Kingdom, Germany, France, Spain, Taiwan, the Netherlands, Australia, the United States, and Sweden), are between 25 and 65 years old, and 44\% are female. 
During app registration, participants self-reported their year of birth and gender (limited to “male” or “female”). 
We acknowledge that this binary classification limits the scope of gender analysis in this study. 
Accordingly, we use the terms “woman” and “female,” as well as “man” and “male,” interchangeably throughout the analysis, while acknowledging that no universally accepted correspondence exists between these terms and an individual’s sex or gender identity~\cite{hawkesworth_sex_2013, ritz_we_2024}.
The home country is determined as the country where the user spent the most time, as indicated by their stop sequences.

\subsection*{Individual mobility networks}\label{sec:meth-netw}
We represent each individual’s one-month sequence of visited locations as a weighted network. Nodes correspond to unique locations, and an edge $i \rightarrow j$ indicates that the individual visited location $j$ immediately after location $i$. The weight $w_{ij}$ of each edge counts the number of such transitions over the observation window (one month). While the network retains a directed structure, we assess network metrics using the corresponding undirected network.

The network representation offers a simple and interpretable mapping between the network structure and individual mobility behavior.
The number of nodes, reflects the repertoire of unique locations visited, $k$, and the sum of edges weights reflects the total activity, $N$--- defined as the sequence length, i.e. the number of visits within the observation window. 
Formally, $N = \sum_{(i,j) \in E} w_{ij} + 1,$ where $E$ is the set of directed edges and $w_{ij}$ their weights. 
The $+1$ accounts for the fact that a sequence of $N$ visits yields $N-1$ transitions. Because all sequences span the same time window, $N$ is directly comparable across individuals.

Notably, the number of visits to a specific node can be approximated by summing the weights of its incident edges and dividing by two, since each visit typically involves both an arrival and a departure. 
This approximation holds for most nodes but fails at the sequence boundaries: the first and last nodes may involve only a single connection, leading to an underestimation of their true visit count. 
For example, in the sequence $[1,2,1,2,1]$, the total activity is $N=5$, which equals the sum of edge weights ($w_{12}+w_{21}=4$) plus one. Node $1$ is visited three times, not twice as predicted by the degree-based approximation, while node $2$ is visited exactly twice.
In practice, these discrepancies are minor, and node visit counts remain approximately proportional to their degree divided by two.

\subsection*{Country-balanced estimation}\label{sec:meth-imbalance}
Our dataset includes users from multiple countries with unbalanced sample sizes (e.g., 41\% from Japan, 2\% from the United States). 
To avoid results being dominated by countries with larger samples, we compute point estimates by bootstrapping within each country and then averaging across countries, giving equal weight to each. 
Detailed country-level estimates are reported in the Supplementary Materials.

For statistical significance tests (e.g., permutation tests comparing observed values to null distributions), we report the pooled estimates in the main text to maintain comparability with the randomized baseline. 
Robustness is assessed by repeating the same tests separately for each country, with the corresponding results provided in the Supplementary Materials.

\subsection*{One-sided Kolmogorov–Smirnov statistic}\label{sec:meth-KSstat}
To assess if the gender differences in activity and repertoire size are statistically significant, we resort to one-sided Kolmogorov-Smirnov (KS) statistic. 

The one-sided KS statistic for two cumulative distribution functions $M(x)$ and $F(x)$ is defined for $D^{+}$ or $D^{-}$ as
\[
D^{+}=\sup_{x}(M(x)-F(x)),
\]
\[
D^{-}=\sup_{x}(F(x)-M(x)),
\]
depending on the direction of the alternative hypothesis.
Here, $sup_x$ denotes the supremum (the largest value observed difference across all $x$). 
Intuitively, the statistic captures the largest directional difference between the two distribution functions, rather than the absolute difference.

We define $KS_{M>F}$ as the maximum difference where the cumulative distribution of males exceeds that of females ($D^{+}$), and $KS_{M<F}$ as the maximum difference where the cumulative distribution of females exceeds that of males ($D^{-}$).
To assess significance, we compare the observed values to a null distribution of the statistics obtained through random permutation of the gender labels.

Having verified that the distributions do not cross more than two times (Fig.\ref{fig:1}c), we distinguish two cases: 
(i) one distribution has significantly more mass in one tail (only one KS statistic is significant), or 
(ii) one distribution is significantly wider (both KS statistics are significant). 
A detailed account of these scenarios can be found in the Supplementary Material~\ref{sisec-KS}.

\subsection*{Robust Coefficient of Variation}\label{sec:meth-rcv}
The coefficient of variation (CV) is a standard measure of relative dispersion, but it is highly sensitive to outliers and skewed distributions. 
To overcome these limitations, we use the Robust Coefficient of Variation (RCV)\cite{arachchige_robust_2020}, which replaces the mean and standard deviation with the median and median absolute deviation (MAD). The MAD is
\[
\text{MAD} = \text{median}\{|x_i - m|\},
\]
with \(m\) the sample median. To ensure comparability with the standard deviation under normality, the MAD is standardized by the constant \(1.4826 = 1/\Phi^{-1}(3/4)\). The RCV is then
\[
\text{RCV} = \frac{1.4826 \times \text{MAD}}{m}.
\]
When bootstrapping RCV values, we account for the unbalanced sample sizes between males and females by subsampling males to match the number of females. This ensures that differences in RCV are not driven by unequal group sizes.

\subsection*{Network metrics}\label{sec:meth-nwmet}
We quantify structural properties of individual mobility networks using three standard metrics.
First, we compute the \emph{average clustering coefficient}, which measures the tendency of visited locations to form tightly connected subgroups. For a node $i$ with degree $k_i$, the local clustering coefficient is
\[
C_i = \frac{2e_i}{k_i(k_i - 1)},
\]
where $e_i$ is the number of edges between the neighbors of $i$. The network-level clustering coefficient is then obtained as
\[
C = \frac{1}{|V|} \sum_{i \in V} C_i,
\]
where $|V|$ denotes the number of nodes.  

Second, we measure the \emph{number of cycles}, defined as the count of simple closed paths in the network (e.g., A--B--C--A). 
Cycles capture recurrent multi-stop patterns beyond back-and-forth travel.  

Finally, we assess \emph{degree centrality}, which quantifies the relative importance of each node. The degree centrality of a node $i$ is defined as
\[
d_i = \frac{k_i}{|V|-1},
\]
where $k_i$ is the degree of node $i$. We compute degree centrality across all nodes and report values for the top one, two, and three most central nodes.
We discuss the relation between these nodes and the home/work nodes in Supplementary Material~\ref{sisec:1degect-Home}.

We validate our results against additional network metrics: transitivity, network density, average cycle length, and network diameter (see Supplementary Material~\ref{sisec:match-extnwmet}).

\subsection*{Home and Work Detection}\label{sec:meth-howde}
We detect home and work locations using \textit{HoWDe}, a validated algorithm for longitudinal GPS data~\cite{desojo_howde_2025}.
HoWDe integrates multiple heuristics—including visit duration, temporal regularity, and frequency of occurrence—to identify Home and Work locations.
The algorithm is designed to be robust to irregular travel behavior and data sparsity. Validation against ground-truth datasets demonstrates high accuracy across demographic groups, settlement types, and countries~\cite{desojo_howde_2026}.

\subsection*{Nearest neighbor matching}\label{sec:meth-matching}
To disentangle structural network differences from those driven by baseline activity and repertoire size, we apply nearest-neighbor matching~\cite{stuart_matching_2010}.  
Each female user is matched to the most similar male user in terms of (i) total number of visits, $N$, and (ii) repertoire size, $k$.
Prior to matching, both covariates are standardized. 
We then use a $k$-nearest neighbors algorithm with $k=1$ and Euclidean distance as the metric. This ensures that each female is paired with her single closest male counterpart in covariate space.  

We conduct this exercise separately for all users combined and within subgroups defined by activity level (inactive, moderate, active). 
To ensure comparability, we constrain the allowable difference in repertoire size between matched pairs. For the full population, we require exact matching in repertoire size ($\Delta k \leq 1$).
Within activity subgroups, we allow small tolerances reflecting group-specific variability: a maximum of one location for inactive users, two for moderate users, and four for active users. These thresholds correspond to approximately the 10th percentile of repertoire size in each group.  

After matching, we compute the average of the difference in network metrics ($X$) between male and female pairs ($X_{Male}$, $X_{Female}$) relative to their mean: $\frac{X_{Male}-X_{Female}}{(X_{Male} + X_{Female})/2}$. 
To evaluate significance, we compare these values to a null distribution obtained by randomly permuting gender labels while preserving the matched pipeline. 
This randomization procedure provides a baseline against which the observed gender differences can be evaluated.

\subsection*{Tour identification}\label{sec:meth-tours}
We identified \emph{tours} from each individual’s monthly sequence of visited locations by iterating through the ordered list of stops. 
A tour was defined as a sequence of stops that begins and ends to the same anchor location. A tour ends upon the first return to that same location.

For each identified tour, we recorded: the ordered list of places visited, the cumulative distance traveled between consecutive locations (\textit{tour cost}), the round-trip distance from each visited place to the tour’s origin (\textit{tour reward}), the round-trip distances from the most distant location to the tour’s origin (\textit{round-trip distance to the furthest location}), and the time between departure from and return to the origin (\textit{the total duration of the tour}). Distances were computed as great-circle distances between geographic coordinates, and durations were derived from arrival and departure timestamps.

We further identified anchor locations as departure places where the duration of the associated tour was shorter than the time the individual had spent at that location prior to leaving. Anchor locations thus represent anchors of mobility behavior, from which short trips typically originate. We restricted the set of tours to those originating at anchor locations and containing at least two distinct stops, thereby excluding trivial sequences such as consecutive repeated visits to the same location. 

This representation aligns with activity-based approaches in transportation research, which treat tours anchored at habitual locations as fundamental units of travel behavior~\cite{daisy_trip_2020, paleti_joint_2011, primerano_defining_2008}.

When evaluating \textit{tour efficiency}, we focused on the one-month sequences and computed the difference between each sequence’s estimated \textit{reward}—the sum of round-trip distances from home to all visited locations—and its \textit{cost}, given by the total distance traveled, normalized by the \textit{reward}: $1-\frac{\text{cost}}{\text{reward}}$.

\subsection*{Network navigability}\label{sec:meth-nwnavig}
We assess the efficiency of movement on individual mobility networks, defined as the average ease of reaching any node from any other. To this end, we consider two complementary routing models widely used in the network science literature: (i) \emph{random diffusion}, which captures exploratory movement via stochastic routing, and (ii) \emph{shortest-path navigation}, which represents optimal routing strategies.

\subsubsection*{Mean First Passage Time (MFPT)}\label{sec:meth-mfpt}
To quantify navigability under random diffusion, we estimate the \emph{mean first passage time} (MFPT) across network nodes.  
For a source node $i$ and a target node $j$, the first passage time $H_{ij}$ is the number of steps a random walker requires to reach $j$ for the first time.  
The MFPT is then defined as the expected value of this random variable, $\langle H_{ij} \rangle = \mathbb{E}[H_{ij}]$,
averaged over multiple random walk realizations~\cite{lovasz_random_1993, newman_networks_2016}.

We compute MFPT analytically using the spectral properties of the normalized adjacency matrix, following the formulation of Lovász~\cite{lovasz_random_1993}. 
For each user network, we construct an undirected weighted adjacency matrix $A$ and the corresponding degree matrix $D$. 
We then form the normalized adjacency matrix $N = D^{-\frac{1}{2}} A D^{-\frac{1}{2}}$.

Let $\{\lambda_n, w_n\}_{n=1}^{|V|}$ denote the eigenvalues and orthonormal eigenvectors of $N$, ordered such that 
\[
1 = \lambda_1 > \lambda_2 \ge \cdots \ge \lambda_{|V|}.
\]
The eigenvector $w_1$ associated with $\lambda_1 = 1$ corresponds to the stationary distribution and is excluded from the hitting-time expansion.

Lovász showed that the MFPT from node $i$ to node $j$ admits the closed-form expression
\[
H_{ij} 
= 2|E|
\sum_{n=2}^{|V|}
\frac{1}{1 - \lambda_n}
\left(
\frac{w_{n,j}^2}{d_j}
-
\frac{w_{n,i}\, w_{n,j}}{\sqrt{d_i d_j}}
\right),
\]
where $|E|$ is the number of edges, and $d_i$ and $d_j$ are the degrees of nodes $i$ and $j$, respectively. 
This computation yields MFPTs for all ordered source--target node pairs once the spectral decomposition of $N$ is available.

From these pairwise quantities, we compute two network-level measures.  
First, the \emph{network-wide MFPT}, obtained by averaging $\langle H_{ij} \rangle$ across all ordered source–target pairs $(i,j)$ in a user’s network.  
This metric captures the overall ease of reaching any location from any other.  
Second, the \emph{global MFPT (GMFPT)} of the home node, defined as the MFPT to the home location when averaging over all possible starting nodes $i \neq \text{home}$.  
This quantity reflects the anchoring role of home in mobility networks.

\subsubsection*{Global efficiency}
To quantify navigability under optimal routing, we compute the \emph{global efficiency} of each mobility network.  
For a pair of nodes $i$ and $j$, the efficiency is defined as the inverse of their shortest-path distance,  
\[
\epsilon_{ij} = \frac{1}{d_{ij}},
\]
where $d_{ij}$ is the topological shortest path length. Global efficiency is then the mean efficiency across all ordered node pairs~\cite{latora_efficient_2001},
\[
E_{\mathrm{glob}} = \frac{1}{N(N-1)} \sum_{i \neq j} \frac{1}{d_{ij}},
\]
with $\epsilon_{ij}=0$ when no path exists. Values range between 0 and 1, reaching the maximum in a fully connected network.  

To incorporate spatial constraints, we compute a distance-weighted variant following~\cite{vragovic_efficiency_2005}. Let $\tilde{d}_{ij}$ denote the length of the shortest physical path between $i$ and $j$, obtained as the sum of edge distances, and let $l_{ij}$ be their direct Euclidean distance. The weighted efficiency is defined as
\[
E_{\mathrm{glob}}^{\mathrm{dist}} = \frac{1}{N(N-1)} \sum_{i \neq j} \frac{l_{ij}}{\tilde{d}_{ij}}.
\]
This measure evaluates how closely mobility networks approximate direct connections in physical space, recognizing that paths composed of multiple links can remain nearly as efficient as direct routes.

\section*{Code and data availability}
Data required to ensure replicability of the analyses conducted in this study are available at: \href{https://doi.org/10.11583/DTU.31835038}{10.11583/DTU.31835038}. 
Code to reproduce all analyses is openly available at :\\ \href{https://github.com/sdesojo/gender_mobility_networks}{github.com/sdesojo/gender\_mobility\_networks}. 
All data processing and analysis were carried out in compliance with the European Union’s General Data Protection Regulation (GDPR; Regulation 2016/679).


\section*{Author contributions}
Conceptualization: SDS, SL, LA.
Methodology: SDS, LA.
Investigation: SDS.
Visualization: SDS.
Funding acquisition: LA.
Project administration: LA.
Supervision: LA, SL.
Writing – original draft: SDS, LA.
Writing – review \& editing: SDS, SL, LA.

\section*{Competing interests} Authors declare that they have no competing interests.

\subsection*{Acknowledgments}
The authors thank Michele Coscia for guidance on random-diffusion network metrics. 
We also thank Andrea Baronchelli and Paolo Santi for their thoughtful feedback and careful reading of earlier versions of the manuscript.


\newpage
\bibliographystyle{unsrt}
\bibliography{gender_efficiency}
\clearpage

\begin{center}
    {\LARGE \bfseries Supplementary Materials\\}
    \vspace{1cm}
    {\Large Women’s mobility networks enable more efficient travel\\}
    \vspace{0.5cm}
    {\large Sílvia de Sojo, Sune Lehmann, Laura Alessandretti}
\end{center}

\setcounter{figure}{0}
\renewcommand{\thefigure}{S\arabic{figure}}

\renewcommand*\contentsname{List of Supplementary Contents}  
\tableofcontents


\clearpage

\section{Country-level gender differences in median activity and repertoire size}\label{sisec:bsctry}

\begin{figure}[H]
\centering
\includegraphics[width=1\linewidth]{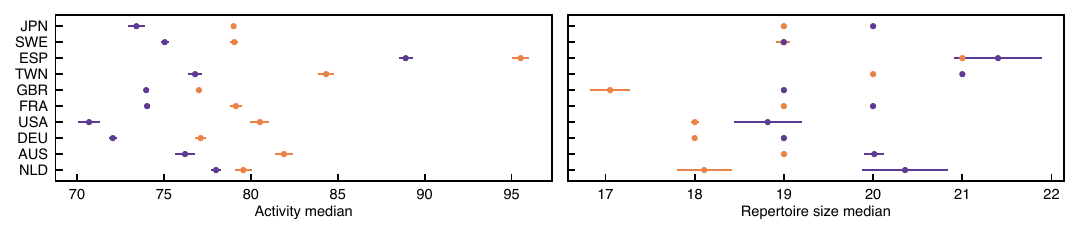}
\caption[Country-level bootstrapped medians of activity and repertoire size.]{\textbf{Country-level bootstrapped medians of activity and repertoire size}
Mean and standard error of the bootstrapped medians (see Methods) for activity (left) and repertoire size (right, shown separately for men (orange) and women (purple).}
\label{fig:si_bsmed_ctry}
\end{figure}

We report country-level bootstrapped medians of activity (monthly visit count) and repertoire size (number of unique locations visited) for men and women (see~figure~\ref{fig:si_bsmed_ctry}). 
Results are shown as the mean and standard error across bootstrap replicates (see Materials and Methods). 

Men exhibit higher activity than women across all countries. Women, however, consistently visit a broader set of unique locations except in Sweden and Spain, where the difference is not statistically significant. 

\begin{figure}[H]
\centering
\includegraphics[width=1\linewidth]{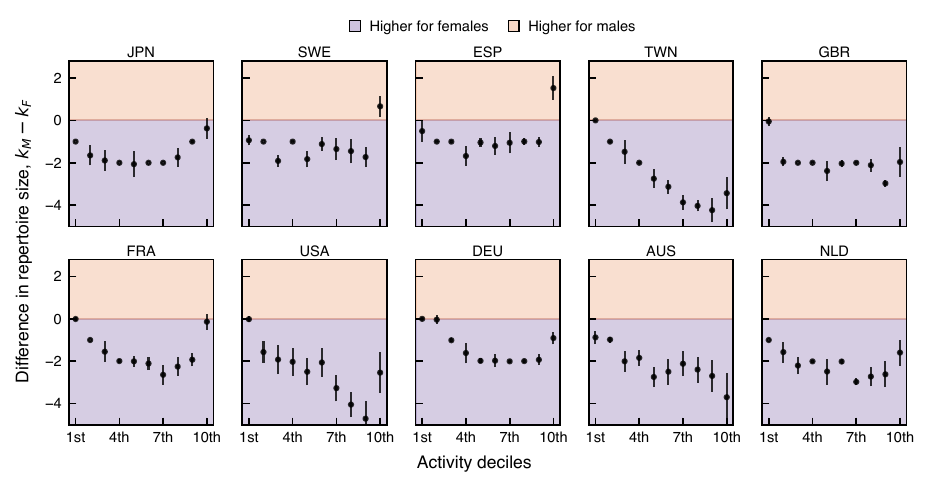}
\caption[Country-level gender differences in repertoire size by activity quantile]{\textbf{Country-level gender differences in repertoire size by activity quantile.}
Gender differences in the median repertoire size, stratified by activity deciles and shown separately for each country.
Error bars indicate the standard error of the difference in medians, estimated via country-balanced bootstrapping (see Methods).
Background shading highlights where repertoires are larger for men (light orange) or for women (light purple).}
\label{fig:si_Nkquant_ctry}
\end{figure}

This pattern holds across the activity distribution: when stratified by activity decile, women maintain larger repertoires than men in every decile and country (see~figure~\ref{fig:si_Nkquant_ctry}). Together, these findings confirm that the gender differences in median repertoire size identified in the main analysis replicate consistently across national contexts.

\section{One-sided Kolmogorov–Smirnov statistics}\label{sisec-KS}
\subsection{Interpreting one-sided Kolmogorov–Smirnov statistics}
As detailed in the Methods, we use the one-sided Kolmogorov–Smirnov (KS) statistic to evaluate the significance of gender differences in activity, repertoire size, and later in network metrics.
To assess the direction of disparities, we employ one-sided tests.

One-sided KS statistics can capture two distinct scenarios:
(i) one distribution has significantly greater mass in a single tail, leading only one KS statistic to be significant; or
(ii) one distribution is significantly wider, in which case both one-sided KS statistics become significant.

These cases highlight that the KS framework is sensitive not only to shifts in central tendency but also to differences in distributional spread.

\begin{figure}[H]
\centering
\includegraphics[width=1\linewidth]{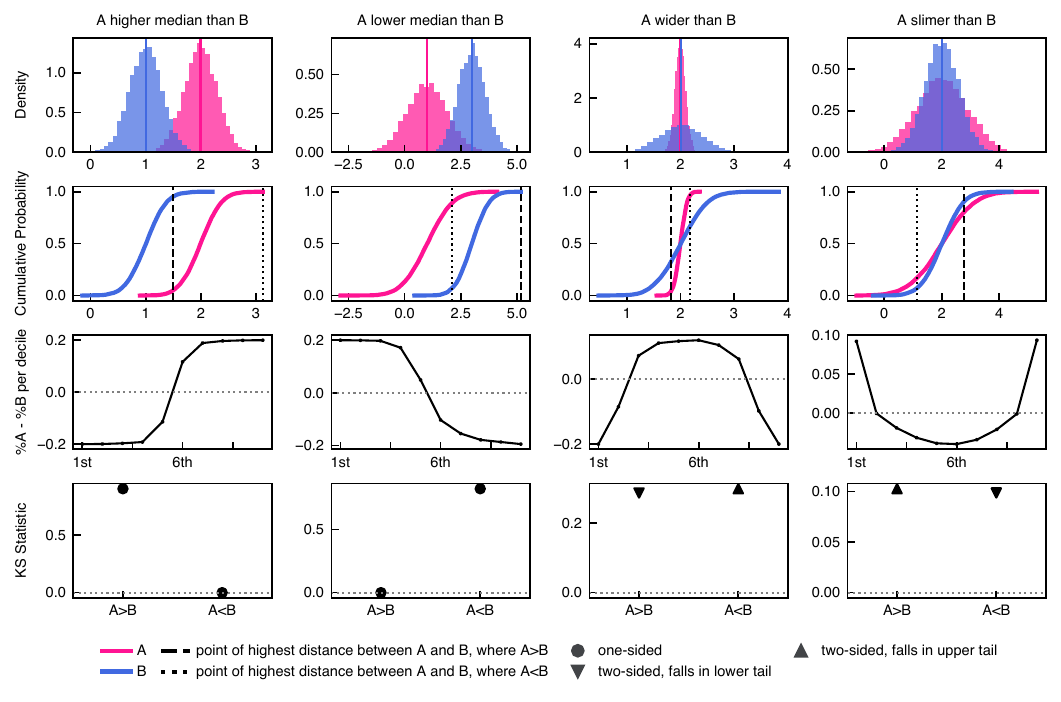}
\caption{\textbf{Interpreting one-sided KS tests.}
Illustrative examples of distinct distributional scenarios and their corresponding KS test outcomes.  
Top panels show the distributions and their medians.  
The second row displays the cumulative probability distributions; vertical lines mark the point of maximum difference, with dashed lines indicating the point for the KS test $A>B$ and dotted lines for $A<B$.  
The third row shows decile-based comparisons of $A-B$, highlighting which parts of the distribution drive the largest differences.  
Bottom panels depict the resulting one-sided and two-sided KS statistics: case (i), when only one test is significant (circles), and case (ii), when both are significant (triangles).  
Upward-pointing triangles denote significance falls in the upper tail of the distribution, while downward-pointing triangles denote significance falls in the lower tail.}
\label{fig:KS_interpet}
\end{figure}

Figure~\ref{fig:KS_interpet} illustrates these scenarios. The first two columns correspond to case (i), where distributions differ primarily in their median. The last two columns correspond to case (ii), where distributions differ in spread.

\begin{figure}[H]
\centering
\includegraphics[width=1\linewidth]{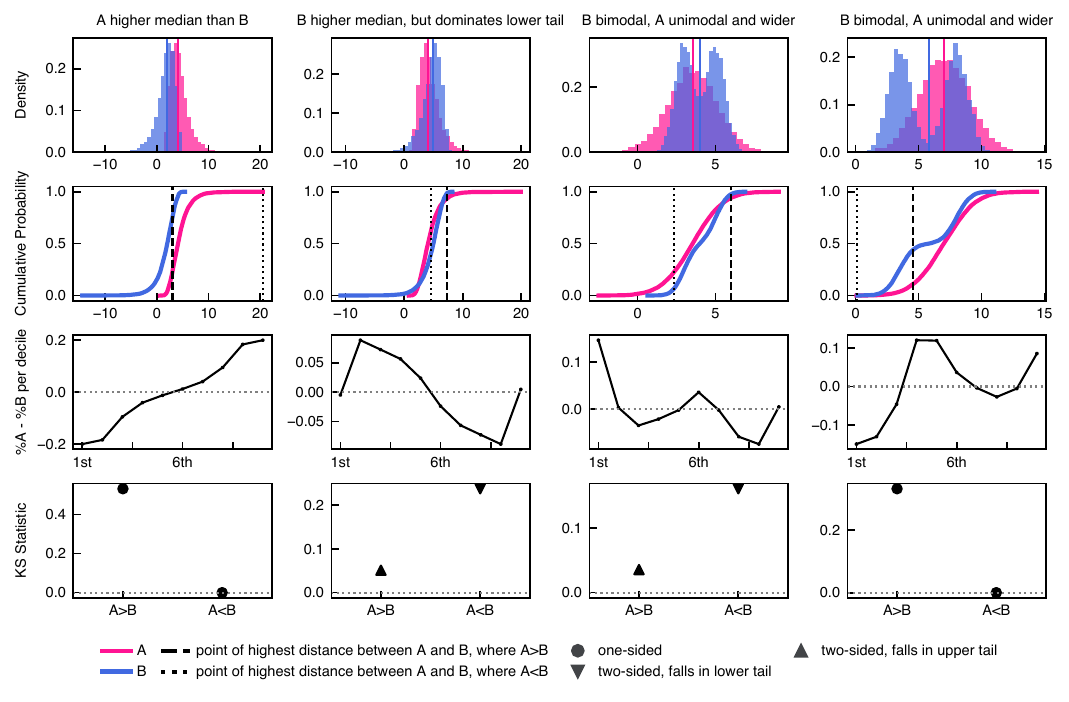}
\caption{\textbf{Limitations of one-sided KS tests.}
Illustrative scenarios where the KS statistic may be misleading.
Panels follow the same layout as Fig.~\ref{fig:KS_interpet}.
We detail below the cases corresponding to each column:
\textbf{Case 1:} distributions cross only once, and the KS statistic reliably captures the difference.
\textbf{Case 2:} the distribution with the higher median also dominates the lower tail, resulting in multiple crossings and potentially leading to the misinterpretation that $B$ dominates the extremes.
\textbf{Cases 3–4:} bimodal distributions for $B$ generate multiple crossings, which distort the KS interpretation.}
\label{fig:KS_interpet_issues}
\end{figure}

In these examples, distributions cross either once (i) or at most twice (ii), conditions under which the KS statistic reliably captures tail differences.
However, when distributions cross three or more times, the KS statistic may fail to provide a faithful characterization.
This occurs, for example, in bimodal distributions (see columns 3 and 4 in Fig.\ref{fig:KS_interpet_issues}) or when the distribution with the higher median also dominates the lower tail (see column 2 in Fig.\ref{fig:KS_interpet_issues}).
Because the KS statistic reflects only the single point of maximum distance between two distributions, it may overlook additional disparities elsewhere.

Overall, one-sided KS tests provide a robust and interpretable framework when distributions differ primarily in central tendency or spread, but they are less reliable in more complex cases involving multiple crossings.

\subsection{Country-level significance of gender differences in activity and repertoire size from one-sided KS tests}

\begin{figure}[H]
\centering
\includegraphics[width=1\linewidth]{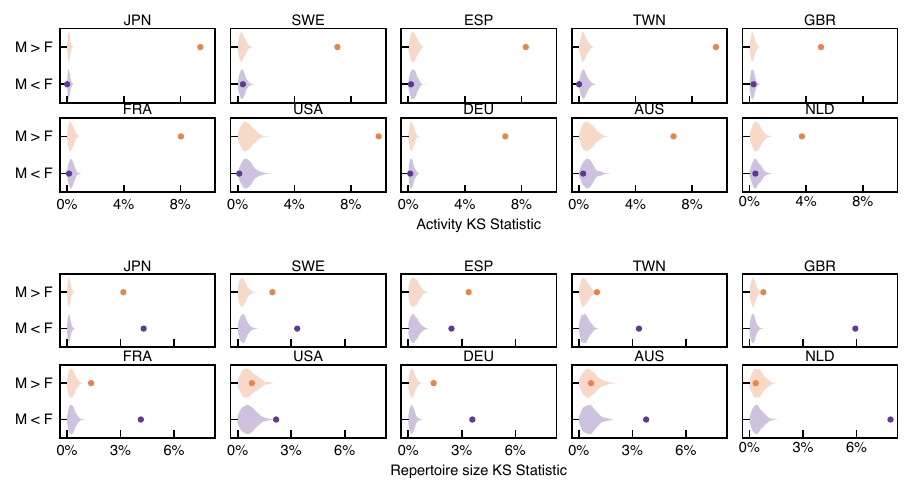}
\caption{\textbf{Country-level one-sided KS tests of gender differences in activity and repertoire size.}
\textbf{Left:} Illustrative KS outcomes when one distribution consistently dominates (top) versus when one distribution is broader (bottom).
\textbf{Right:} KS statistics for the hypotheses $M>F$ (orange background) and $M<F$ (purple background), shown separately by country. Grey violin plots represent the reference distribution obtained from random permutation of gender labels.}
\label{fig:si_KS_ctry}
\end{figure}

Here, we present the KS results stratified by country (see figure~\ref{fig:si_KS_ctry}).
Since the male and female distributions cross only once or twice (see main text, Fig.1.c), the KS statistic provides a fair basis for comparison(see previous section).

Across countries, one-sided KS tests consistently uncover significant gender disparities in both activity and repertoire size.
Men are more active than women in all countries.
In terms of repertoire size, men tend to dominate the extremes of the distribution, with the notable exceptions of the United States, Australia, and the Netherlands, where women exhibit broader repertoires.

This expanded analysis illustrates how one-sided KS statistics capture both consistent and context-specific gender differences, distinguishing shifts in central tendency from broader distributional spread.

\section{Stratification of users by activity and repertoire size}\label{sisec-Nkstrat}
To address heterogeneities in activity ($N$) and repertoire size ($k$), we stratify individuals into three groups based on their joint deciles:  
\textit{inactive} (1st–3rd deciles), \textit{moderate} (4th–7th deciles), and \textit{active} (8th decile and above). 

\begin{figure}[H]
    \centering
    \includegraphics[width=0.9\linewidth]{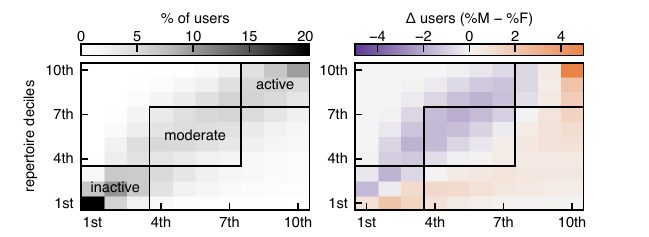}
    \caption{\textbf{Activity–repertoire stratification of users.}
    \textbf{Left:} Percentage of users in each bin defined by the decile of activity and repertoire size. Overlaid boxes define three mobility groups: inactive (low activity and repertoire), moderate, and active (high activity and repertoire).
    \textbf{Right:} Difference in user composition by gender for each bin, calculated as the percentage point difference between male and female users ($\%M-\%F$). Men are overrepresented in the most extreme bins—those with either very low or very high activity and repertoire.}
    \label{fig:si-frac-AMI-MF}
\end{figure}

Fig.~\ref{fig:si-frac-AMI-MF} shows the resulting partition (left) and the gender composition across bins (right). Men are disproportionately represented in the Inactive and Active groups, while women are more frequently found in the Moderately Active group. Because $43.2\%$ of users fall outside these three subgroups, we also report results for the full sample.

\section{Gendered structural differences across extended network metrics}\label{sisec:match-extnwmet}

We expand the analysis presented in Section~“Structural differences in gendered mobility networks” by examining a broader set of network descriptors.
Using the same nearest-neighbor matching procedure described in the Methods, we measure gender differences across the following additional metrics:

\begin{itemize}
    \item \textbf{Transitivity}: measures the fraction of closed triangles among all possible triplets in the network. Possible triangles are identified through triads (two edges sharing a vertex). Formally, transitivity is computed as $3 \times \text{number of triangles} / \text{number of triads}$.
    
    \item \textbf{Density}: the fraction of realized links out of all possible links, indicating how saturated a mobility network is. Density takes the value 0 for an edgeless graph and 1 for a complete graph.
    
    \item \textbf{Average node connectivity}: quantifies the robustness of a network by measuring, on average, how many nodes must be removed to disconnect two arbitrary nodes. It is computed as the mean \textit{local node connectivity} across all node pairs, where local node connectivity is the minimum number of nodes whose removal disconnects the pair. Higher values indicate more robust and well-connected networks.
    
    \item \textbf{Diameter}: the length of the longest shortest path between any two nodes, providing a measure of the network’s overall extent.

\end{itemize}

\begin{figure}[H]
    \centering
    \includegraphics[width=1\linewidth]{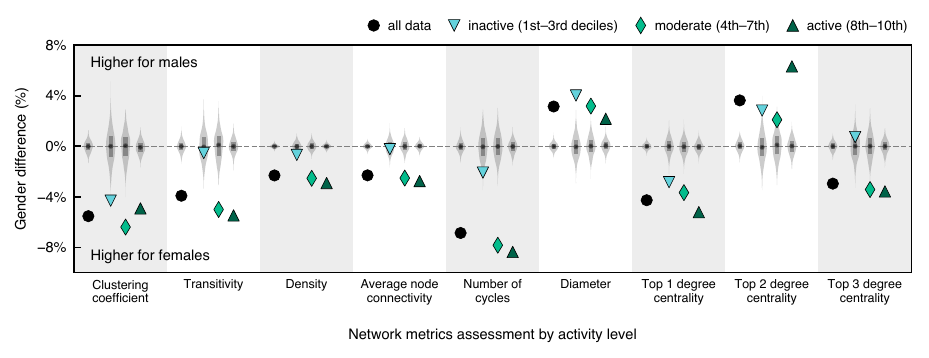}
    \caption{\textbf{Nearest-neighbor matching for additional network metrics.}
    Relative gender difference between males and females, measured as $\frac{Males-Females}{(Males+Females)/2}$, for different network quantities (from left to right): clustering, transitivity, density, average node connectivity, number of cycles, diameter, and degree of the first, second, and third highest-degree nodes. To control for confounding factors, results are computed from male–female pairs matched by activity and repertoire size (see Methods). 
    For each metric, results are shown for the following groups: inactive (light blue downward-pointing triangles), moderately active (green diamonds), active (dark green upward-pointing triangles), and all individuals (black circles). Grey violins represent the reference distribution obtained from 1,000 random shuffles of gender labels. 
    }
    \label{fig:si_knn_allmet}
\end{figure}

As shown in Fig.~\ref{fig:si_knn_allmet}, women’s networks consistently exhibit higher values of density, transitivity, and average node connectivity, as well as smaller diameters, compared to men’s networks.
These results reinforce the main finding that structural differences in mobility networks are systematic across multiple descriptors.

\section{Degree centrality as a proxy for home and work locations}\label{sisec:1degect-Home}
The degree centrality of nodes in individual mobility networks offers a natural proxy for the role of home and work. 
We validate this assumption by comparing the top one and top two most central nodes with home and work locations detected using the state-of-the-art \textit{HoWDe} algorithm (see Methods).
The most central node corresponds to the home in $92.01\%$ of users with a detected home, while the second most central node corresponds to the workplace in $54.09\%$ of users.
These findings support the interpretation of degree centrality as a reliable indicator of anchor locations in individual mobility networks.

\section{Country-level structural differences in gendered mobility networks}\label{sisec:match-nwmetCtry}
\begin{figure}[H]
    \centering
    \includegraphics[width=1\linewidth]{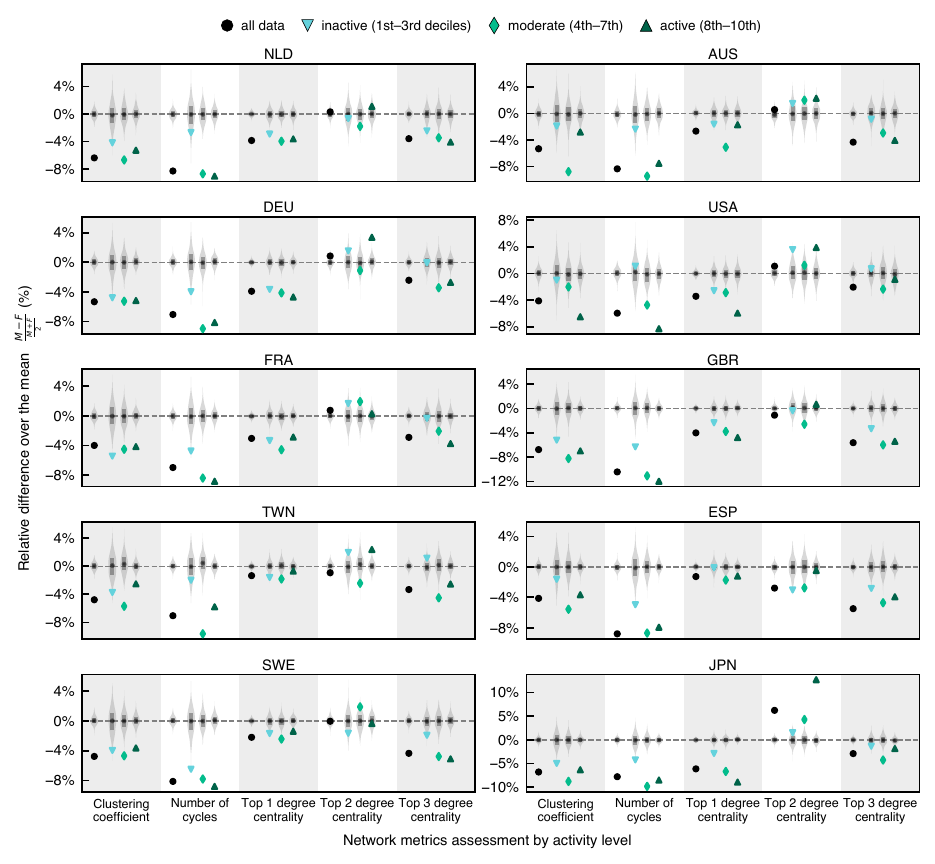}
    \caption{\textbf{Country-level nearest-neighbor matching.} 
    Each subplot represents one country. Relative gender difference between males and females, measured as $\frac{Males-Females}{(Males+Females)/2}$, for different network quantities (from left to right): clustering, number of cycles, and degree of the first, second, and third highest-degree nodes. To control for confounding factors, results are computed from male–female pairs matched by activity and repertoire size (see Methods). 
    For each metric, results are shown for the following groups: inactive (light blue downward-pointing triangles), moderately active (green diamonds), active (dark green upward-pointing triangles), and all individuals (black circles). Grey violins represent the reference distribution obtained from 1,000 random shuffles of gender labels. }
    \label{fig:si_knn_ctryweighted}
\end{figure}

We report country-level results for the network metrics used to assess structural differences in mobility between men and women.
Applying the nearest-neighbor matching procedure within each country, we find that the observed gender differences are overall consistent across contexts (see figure~\ref{fig:si_knn_ctryweighted}). 
Few exceptions appear for the most inactive and active subgroups across countries, particularly for the top 2 and top 3 degree centrality groups.
Importantly, across all metrics, when assessing the entire sample (see figure~\ref{fig:si_knn_ctryweighted}, black dot), the gender differences are significant across all countries. The only exception appears for top 2 degree centrality, where gender differences are not significant for the Netherlands, Australia, France, and Sweden.

This confirms that the structural patterns identified in the main analysis are robust across countries and not driven by a particular country sample.

\section{One-sided KS tests across network metrics}\label{sisec:KS-nwmet}
To evaluate gender disparities beyond average differences in network metrics, we apply one-sided Kolmogorov–Smirnov (KS) tests (see Methods).
Because most network metrics depend strongly on repertoire size ($k$, number of nodes) and activity level ($N$, weighted links), we assess significance separately within three activity groups: inactive (1st–3rd deciles), moderate (4th–7th), and active (8th–10th).

Overall, the distributional results are consistent with those obtained using nearest-neighbor matching.
Women tend to display more connected networks, with a greater number of cycles and stronger home anchoring.
For the most active users, however, men exhibit more extreme behavior in clustering, dominating the upper tail of the distribution.

We only identify a discrepancy among inactive individuals, where men show a higher average clustering than women.
Yet even within activity-based groups, substantial heterogeneity remains in $N$ and $k$, suggesting that some differences may be conflated with variation in network size rather than structure alone.
This reinforces that nearest-neighbor matching provides a more precise benchmark for gender differences than distributional tests in isolation.

\begin{figure}[H]
    \centering
    \includegraphics[width=1\linewidth]{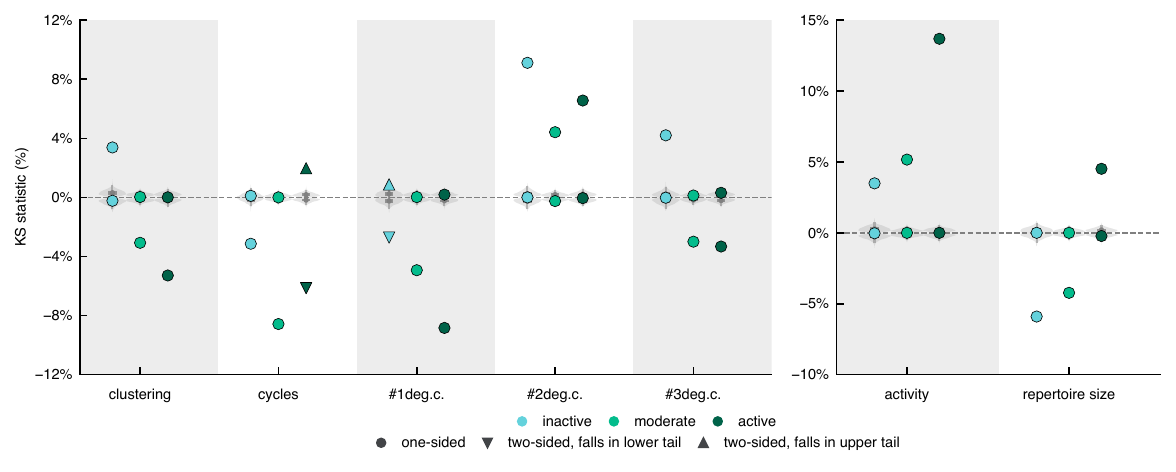}
    
    \caption{\textbf{One-sided Kolmogorov–Smirnov (KS) tests of gender differences in network metrics.} 
    KS statistics are reported for the hypotheses $Males>Females$ (positive values) and $Males<Females$ (negative values), compared against a null distribution generated from 1,000 random shuffles of gender labels (grey violin plots).
    Results are reported for network metrics (left) and mobility metrics (right), and shown separately for inactive (light blue), moderately active (green), and active (dark green) individuals.
    Marker shapes indicate significance: circles denote tests significant for one hypothesis only, while triangles denote significance for both. Downward-pointing triangles correspond to significance in the lower tail of the distribution, and upward-pointing triangles to the upper tail. For a detailed guide on how to interpret the KS tests, see SI Section~\ref{sisec-KS}.}
    \label{si_fig:ks_nw_met}
\end{figure}

\section{Country-level gender differences in tour length}\label{sisec:tourlen-ctry}
To examine whether gendered patterns in tour structure generalise across contexts, we report country-level results for tour length distributions. Across countries and activity levels, men are more likely to rely on simple back-and-forth travel, while two-stop tours are consistently more common among women. 
Longer tours show only minor gender differences across most countries, with the exception of 7–10 stop tours, which are consistently male-predominant. In the particular case of Japan, 3-stop and 11--20-stop tours are also male-predominant.

\begin{figure}[H]
    \centering
    \includegraphics[width=1\linewidth]{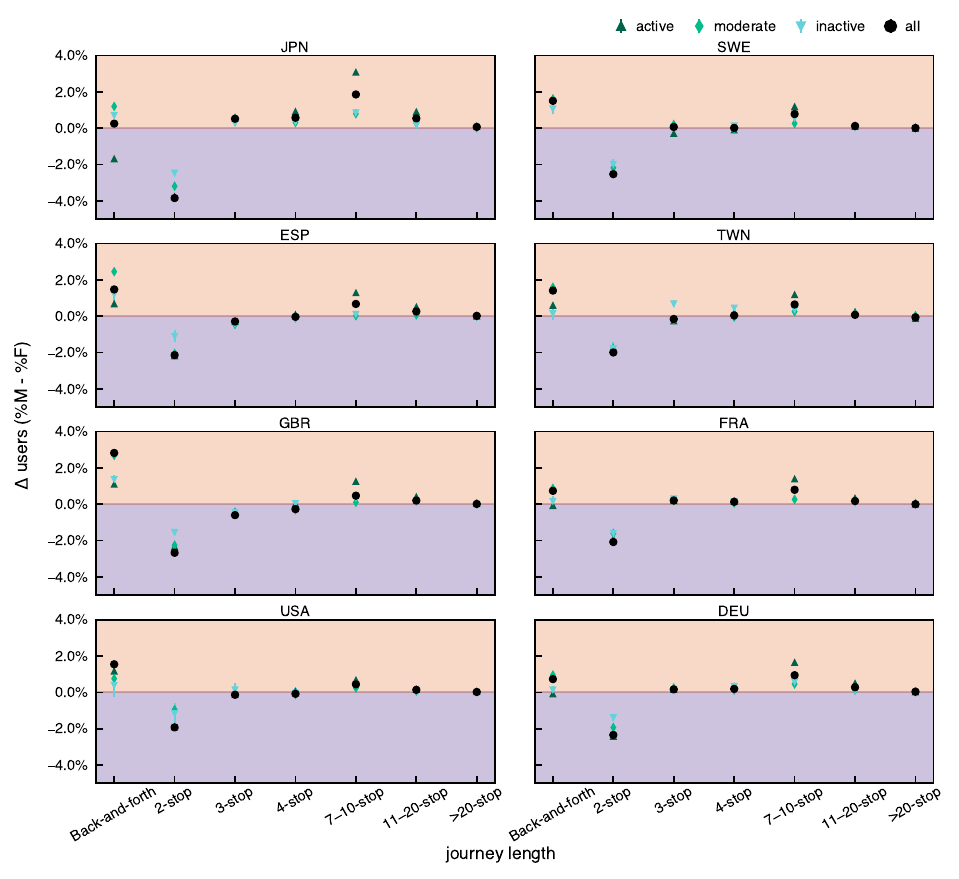}
    \caption{Differences between the share of male and female tours by length, for each country separately (subplots). Marker colors denote activity groups: inactive (light blue downward-pointing triangles), moderately active (green diamonds), active (dark green upward-pointing triangles), and all individuals (black circles). Results show means and bootstrapped standard errors; error bars are in some instances too small to be visible (see Methods).}
    \label{fig:si_tourlen_C}
\end{figure}

\section{Country-level gender differences in tour length by round-trip distance to the furthest location}\label{sisec:tourlen-dist-ctry}

\begin{figure}[H]
    \centering
    \includegraphics[width=1\linewidth]{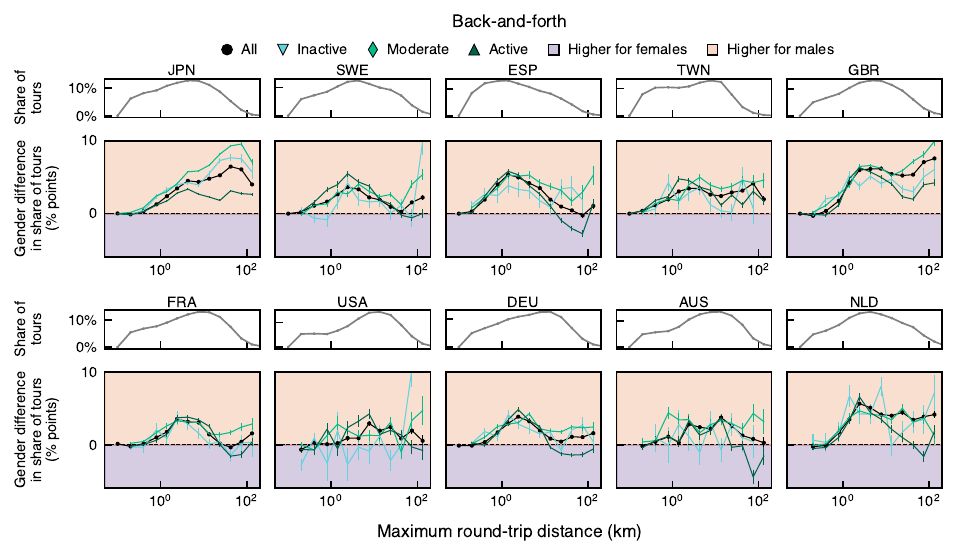}
    \caption{\textbf{Back-and-forth tours by distance bin across countries.} Relative prevalence of back-and-forth tours (A–B–A) by gender, grouped by round-trip distance to the furthest location. Results are shown separately for each country. Marker colors denote activity groups: inactive (light blue downward-pointing triangles), moderately active (green diamonds), active (dark green upward-pointing triangles), and all individuals (black circles). Bootstrapped standard errors are shown, though often too small to be visible (see Methods).}
    \label{fig:tourlen_maxdist_C_BF}
\end{figure}

\begin{figure}[H]
    \centering
    \includegraphics[width=1\linewidth]{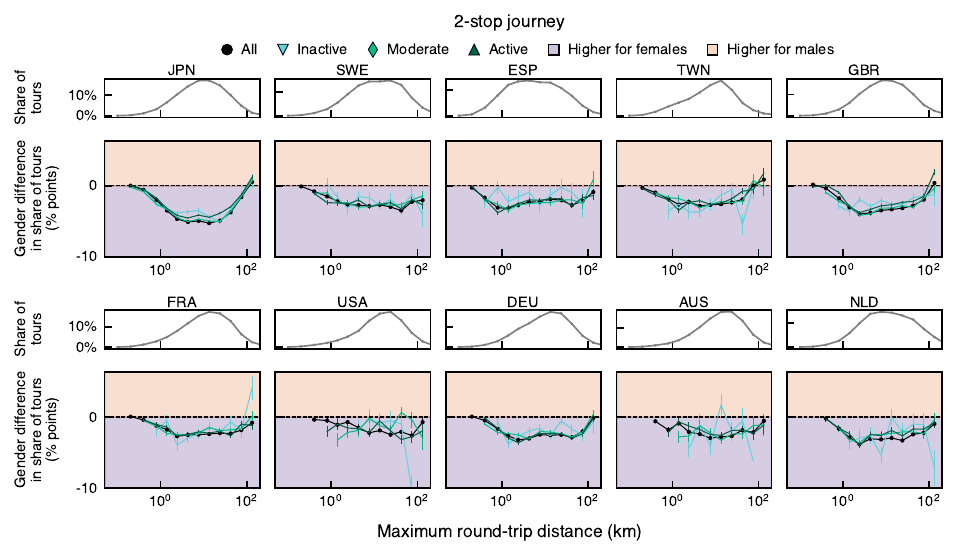}
    \caption{\textbf{Two-stop tours by distance bin across countries.} Relative prevalence of two-stop tours (A–B–C–A) by gender, grouped by round-trip distance to the furthest location. Results are shown separately for each country, with marker colors and error bars as in Fig.~\ref{fig:tourlen_maxdist_C_BF}.}
    \label{fig:tourlen_maxdist_C_2}
\end{figure}

We next examine whether gendered differences in tour structure as a function of round-trip distance to the furthest location are consistent across countries.
We confirm that men are more likely to make back-and-forth tours (see Fig.\ref{fig:tourlen_maxdist_C_BF}), whereas women more often undertake two-stop tours up to distances of about 150 km  (see Fig.\ref{fig:tourlen_maxdist_C_2}).
The prevalence of two-stop tours among women is strikingly consistent across countries.
For back-and-forth tours, some discrepancies appear at longer distances, particularly in Australia and the United States, suggesting that local factors---such as car dependency, urban form, or travel norms---may contribute.

Overall, these findings confirm that the structural differences observed in tour organization are robust across both national contexts and distance levels.

\section{Country-level gender differences in tour length by tour duration}\label{sisec:tourlen-durat-ctry}
We also group tours by their total duration to test whether gendered differences in tour structures show similar results as those observed across maximum round-trip distances.

Consistent with the distance-based analysis, men rely more heavily on back-and-forth tours, while women are more likely to undertake two-stop tours across duration bins up to approximately 12 hours.

For longer tours, gender differences in back-and-forth travel diminish and, in some countries, reverse, whereas two-stop tours remain consistently more common among women.
This suggests that circadian rhythms may play a role: daily cycles of activity and rest could constrain tour organization differently for men and women, particularly over extended durations.

The overall robustness of these results confirms that the greater prevalence of two-stop structures among women cannot be explained by trip-chaining to nearby locations alone, but reflects systematic differences in how tours are organized.

\begin{figure}[H]
    \centering
    \includegraphics[width=1\linewidth]{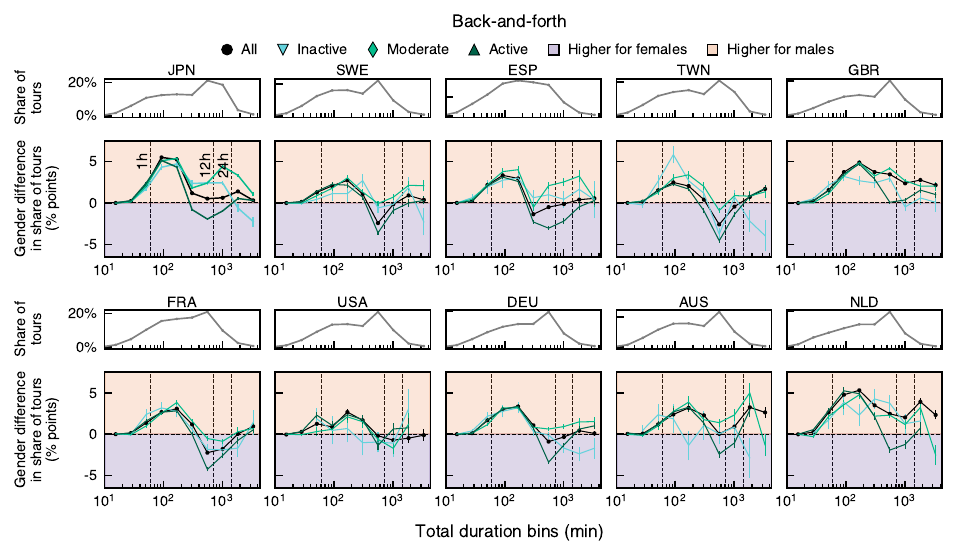}
    \caption{\textbf{Back-and-forth tours by duration bin across countries.} 
    Top: Share of tours by duration bin.
    Bottom: Relative prevalence of back-and-forth tours (A–B–A) by gender, grouped by total tour duration. Results are shown separately for each country (subplots). Marker colors denote activity groups: inactive (light blue downward-pointing triangles), moderately active (green diamonds), active (dark green upward-pointing triangles), and all individuals (black circles). Bootstrapped standard errors are shown, though often too small to be visible (see Methods).}
    \label{fig:si_tourlen_durat_C_BF}
\end{figure}

\begin{figure}[H]
    \centering
    \includegraphics[width=1\linewidth]{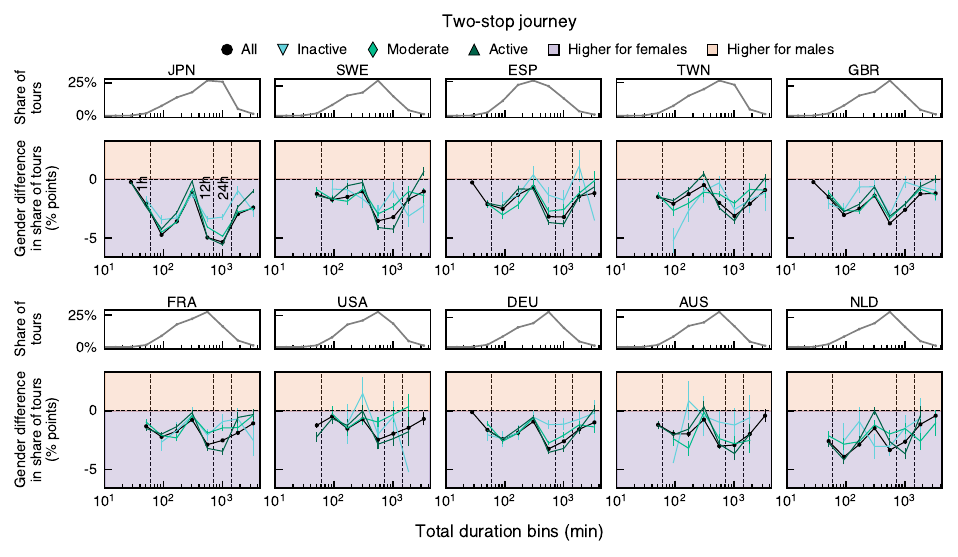}
    \caption{\textbf{Two-stop tours by duration bin across countries.}
    Top: Share of tours by duration bin.
    Bottom: Relative prevalence of 2-stop tours (A–B–C-A) by gender, grouped by total tour duration. Results are shown separately for each country (subplots). Marker colors denote activity groups: inactive (light blue downward-pointing triangles), moderately active (green diamonds), active (dark green upward-pointing triangles), and all individuals (black circles). Bootstrapped standard errors are shown, though sometimes too small to be visible (see Methods)}
    \label{fig:si_tourlen_durat_C_B2}
\end{figure}

\section{Country-level gender differences in MFPT and Global efficiency}\label{sisec:navmet-ctry}

We next test whether gender differences in network efficiency are robust across country contexts. Random-walk dynamics measured with network-wide mean first passage time (MFPT) show that women's networks are easier to navigate than men's across most countries when considering the entire sample. Across activity profiles, however, exceptions emerge: results are non-significant for inactive individuals in Japan and Spain, for active individuals in the Netherlands and Germany, and across most activity subgroups in Sweden, France, Australia, and the USA.

For home-based MFPT, women consistently require shorter passage times across most countries, with exceptions limited to inactive and moderate individuals in Spain, and active individuals in Sweden, Taiwan, and Australia.

Global efficiency metrics indicate that women's networks are generally more efficient across both shortest-path and distance-based measures. Exceptions are limited to active individuals in Taiwan and Spain, and inactive and active individuals in Australia for shortest-path efficiency; for distance-based efficiency, exceptions are confined to active individuals in Taiwan and Australia, and inactive individuals in the USA and Australia.

\begin{figure}[H]
\centering
\includegraphics[width=1\linewidth]{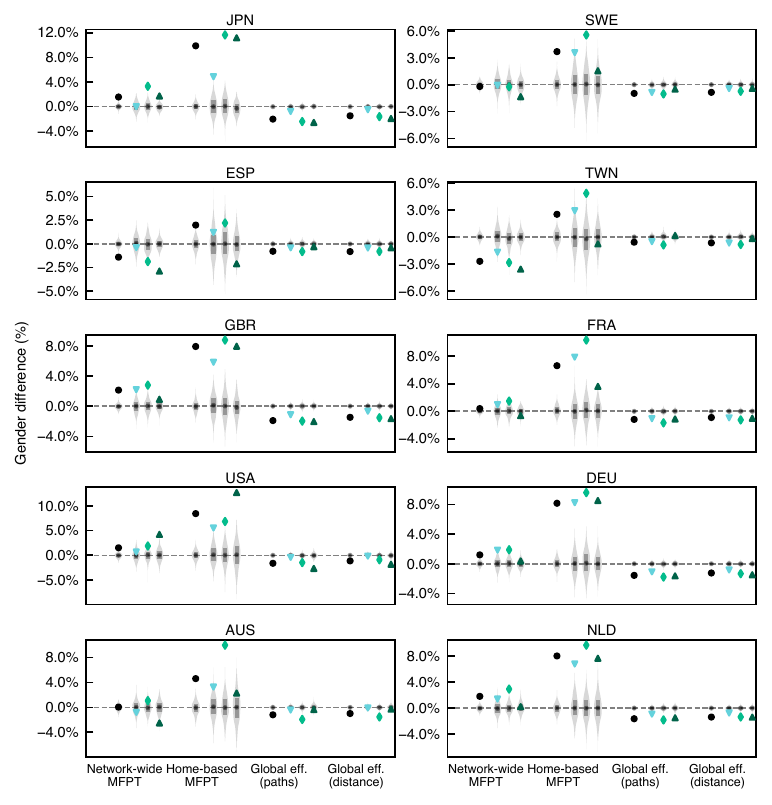}
\caption{\textbf{Country-level gender differences in network efficiency.}
Average relative gender differences across matched pairs, measured as $\frac{X_{Male}-X_{Female}}{(X_{Male} + X_{Female})/2}$, in network-wide and home-based mean first passage time (MFPT), global efficiency (unweighted and weighted by distance).
Averages are computed across male–female pairs matched by activity and repertoire size (see Methods)
Marker colors denote activity groups: inactive (light blue downward-pointing triangles), moderately active (green diamonds), active (dark green upward-pointing triangles), and all individuals (black circles).
Gray violins show reference distributions from 1,000 random shuffles of gender labels.
Negative values indicate higher values for females. 
Bootstrapped standard errors are shown, though often too small to be visible (see Methods).
}
\label{fig:si_navmet_ctry}
\end{figure}

\section{Correlation between travel time and distance}\label{sisec:travel-time}

In this section, we study the empirical relationship between monthly traveled distance and monthly travel time.
Specifically, we compute for each individual the total distance traveled and the total time spent traveling over a month and compare their logarithmic values to reduce the influence of extreme values.

Travel time is estimated as the elapsed time between the start and end of consecutive recorded stop locations.
Because this estimate can be affected by temporal sparsity in the data---for example, long gaps between observed stops may artificially inflate inferred travel times---we restrict the analysis to a subset of users with sufficient temporal coverage.
Specifically, we compute each individual’s median daily share of time with a registered stop and exclude the bottom 10\% of users, retaining only those with a temporal coverage greater than 0.7.

\begin{figure}[H]
    \centering
    \includegraphics[width=0.75\linewidth]{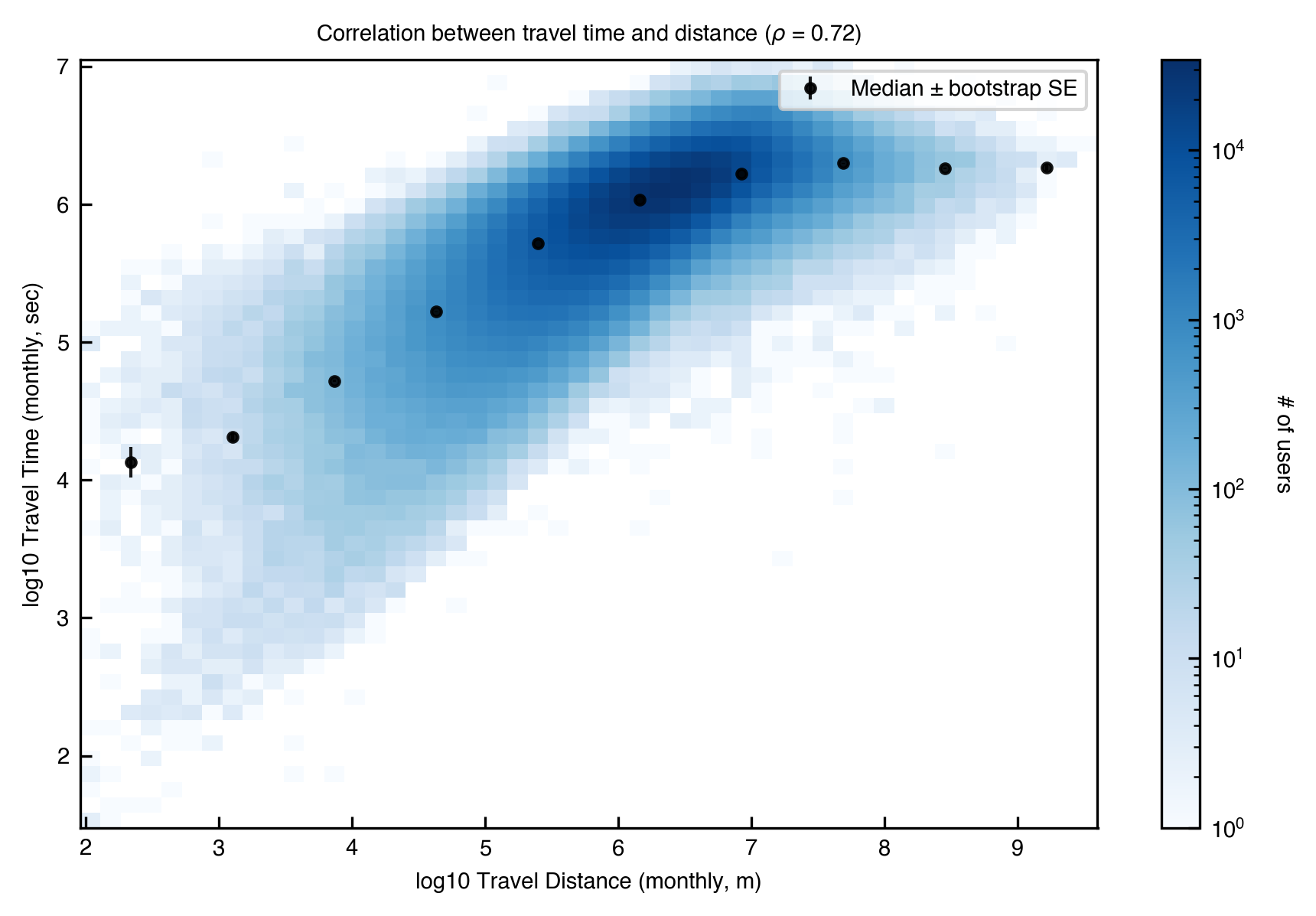}
    \caption{Correlation between monthly travel distance and travel time. Density heatmap of individuals’ total monthly travel distance and travel time (both on a logarithmic scale). Colors indicate the number of users per bin. The strong positive correlation ($\rho_{time, distance}=0.72$) supports the use of travel distance as a proxy for the travel time budget.}
    \label{fig:travel-time-dist}
\end{figure}

Supplementary Fig.~\ref{fig:travel-time-dist} shows a strong positive association between monthly travel distance and monthly travel time, with a Pearson correlation coefficient of $\rho_{time, distance}=0.72$.

Applying a stricter temporal coverage threshold, retaining only individuals with coverage above 0.92 (top 10\% of users), leads to only a marginal increase in the correlation, which remains high at $\rho_{\text{time, distance}} = 0.75$. 

\section{Gender differences in cost conditional on mobility reward}\label{sisec:cost-reward}

\begin{figure}[H]
    \centering
    \includegraphics[width=0.9\linewidth]{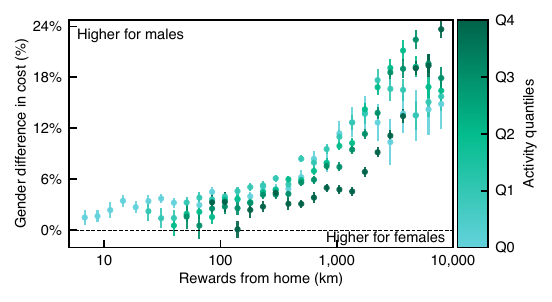}
    \caption{Gender differences in mobility cost compared to achieved reward.
    Relative difference in mobility cost between males and females as a function of achieved mobility reward, stratified by activity quantiles.
    Positive values indicate higher costs for males.
    Across activity levels, males consistently show higher mobility costs than females to achieve comparable rewards, with differences increasing at higher reward levels.}
    \label{fig:si-costreward}
\end{figure}

In this section, we corroborate our analyses on the difference in efficiency across genders by comparing individuals with similar activity level (number of visits) and assessing how much travel distance is required (cost) to reach comparable outcomes (reward).
Supplementary Fig.~\ref{fig:si-costreward} shows the relative gender difference in mobility cost as a function of achieved mobility reward across activity quantiles.
Across nearly all reward levels and activity groups, males consistently exhibit higher mobility costs than females to achieve the same reward.
The magnitude of this difference increases with reward, reaching approximately up to 13-23\% at higher reward levels.

These results indicate that, conditional on achieved reward and overall activity level—approximating a fixed travel time budget—females systematically require less travel to reach comparable outcomes.
This provides evidence that gender differences in tour efficiency reflect differences in trip organization rather than differences in available mobility budgets.

\section{Country-level gender differences in tour efficiency}\label{sisec:eff-ctry}
We next test whether the gender gap in tour efficiency observed in the main analysis is robust across countries.
Tour efficiency is defined as the proportion of distance saved when chaining visits into tours, relative to visiting each location independently (see Methods).
Higher values indicate that individuals structure their mobility more efficiently, minimizing redundant travel.

Across countries, women consistently exhibit higher tour efficiency than men, even when comparing individuals with similar activity levels and sequence rewards.
This confirms that women’s tendency to organize more connected tours translates into systematically greater travel efficiency across national contexts.

\begin{figure}[H]
    \centering
    \includegraphics[width=1\linewidth]{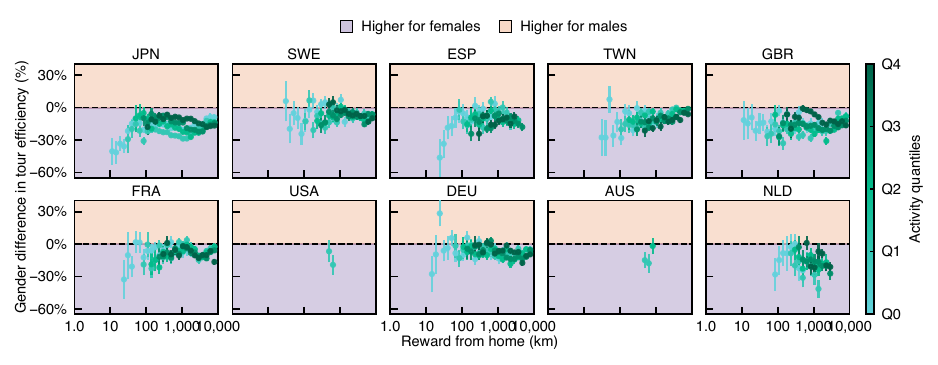}
    \caption{\textbf{Country-level gender differences in tour efficiency.}  
    Relative gender differences, $\frac{Males-Females}{(Males+Females)/2}$, in tour efficiency, computed as $1-\frac{\text{cost}}{\text{reward}}$.  
    Results are grouped by reward (log-transformed distance from home) and activity quantiles.  
    Different shades of grey denote activity quantiles (see colorbar).}
    \label{fig:placeholder}
\end{figure}

\end{document}